\documentclass[aps,prb,preprint,groupedaddress,superscriptaddress, showpacs]{revtex4-1}

\usepackage{graphicx}
\usepackage{bm}
\usepackage{amsfonts}
\usepackage{multirow} 

\draft 

\begin{document}


\title{Photoluminescence studies of Zeeman effect in type-II InSb/InAs nanostructures} 



\author{Ya.V.~Terent'ev}
\email[]{yakov.terentyev@mail.ioffe.ru}
\affiliation{Ioffe Physical Technical Institute of the Russian Academy of Sciences, 26, Polytekhnicheskaya Str., St Petersburg, 194021, Russia}

\author{M.S.~Mukhin}
\affiliation{Ioffe Physical Technical Institute of the Russian Academy of Sciences, 26, Polytekhnicheskaya Str., St Petersburg, 194021, Russia}
\affiliation{Saint Petersburg Academic University —-- Nanotechnology Research and Education Centre of the Russian Academy of Sciences, 8/3, Khlopina Str., St Petersburg, 194021, Russia}

\author{A.A.~Toropov}
\affiliation{Ioffe Physical Technical Institute of the Russian Academy of Sciences, 26, Polytekhnicheskaya Str., St Petersburg, 194021, Russia}

\author{M.O.~Nestoklon}
\affiliation{Ioffe Physical Technical Institute of the Russian Academy of Sciences, 26, Polytekhnicheskaya Str., St Petersburg, 194021, Russia}

\author{B.Ya.~Meltser}
\affiliation{Ioffe Physical Technical Institute of the Russian Academy of Sciences, 26, Polytekhnicheskaya Str., St Petersburg, 194021, Russia}

\author{A.N.~Semenov}
\affiliation{Ioffe Physical Technical Institute of the Russian Academy of Sciences, 26, Polytekhnicheskaya Str., St Petersburg, 194021, Russia}

\author{V.A.~Solov'ev}
\affiliation{Ioffe Physical Technical Institute of the Russian Academy of Sciences, 26, Polytekhnicheskaya Str., St Petersburg, 194021, Russia}

\author{S.V.~Ivanov}
\affiliation{Ioffe Physical Technical Institute of the Russian Academy of Sciences, 26, Polytekhnicheskaya Str., St Petersburg, 194021, Russia}


\date{\today}

\begin{abstract}
Electron spin polarization up to 100\% has been observed in type-II narrow-gap heterostructures with ultrathin InSb insertions in an InAs matrix via investigation of circularly polarized photoluminescence in an external magnetic field applied in Faraday geometry. The polarization degree decreases drastically, changes its sign, and saturates finally at the value of 10\% in the limit of either high temperature or strong excitation. The observed effect is explained in terms of strong Zeeman splitting of the electron conduction band in the InAs matrix and a heavy-hole state confined in the InSb insertion, due to a large intrinsic g-factor of both types of carriers. The hole ground state in a monolayer scale InSb/InAs quantum well, calculated using a tight-binding approach, fits well the observed emission wavelength. Temperature dependence of the emission polarization degree is in good agreement with its theoretical estimation performed in the framework of a proposed phenomenological model.
\end{abstract}


\maketitle 

\section{Introduction}
Important prerequisites for implementation of any semiconductor spintronic device are the functions of spin initiation, spin storage, and spin manipulation. \cite{Dyakonov2008} Individually, these functionalities can be achieved in heterostructures combining some particular semiconductor compounds. Electrical spin initiation, for instance, requires materials with large magnetic g-factor such as II-VI \cite{Fiederling1999} or III-V \cite{Ohno1999} diluted magnetic semiconductors (DMS). Efficient spin storage implies long spin lifetime. This is a natural property of n-type III-V compounds and nanostructures. \cite{Dzhioev2004} To enable spin manipulation by an external electric field strong spin-orbit interaction (SOI) is required as well as high carrier mobility. \cite{Dyakonov2008} Both properties are especially pronounced in III-V narrow-bandgap compounds like e.g. InAs. However, the flexible combination of all these materials in a single high-quality heterostructure is hampered by essentially different growth temperatures, different valencies of the elements involved in the compounds of various chemical groups, as well as by different doping capacities. Hence, fabrication of a single-chip spintronic structure combining different functionalities has yet been a challenge.

In this paper, we propose an approach to this problem based on employment of narrow-bandgap nonmagnetic III-V semiconductors. The narrow-bandgap compounds, such as InSb and InAs, possess the largest values of electron g-factor among known nonmagnetic semiconductors (-50 and -15 for bulk InSb and InAs, respectively). \cite{Pidgeon1967,Ruhle1975} In combination with small effective mass of electrons, high mobility, and strong SOI, this allows one to consider InSb and InAs as  promising components of heterostructures implementing the functions of spin initiation and manipulation. Additionally, heterostructures based on these compounds often demonstrate type-II band alignment providing long lifetime of photoexcited carriers that is an  important advantage in the context of a spin storage functionality. \cite{Hatami1998}

Seeming obvious, this approach, to the best of our knowledge, hasn't been systematically  pursued yet. In particular, the primary issue of  the dependence of the intrinsic electron spin polarization on an external magnetic field, temperature, and density of non-equilibrium carriers has not been addressed in any detail for the heterostructures involving these compounds. The outlined gap in the research can be explained by general non-availability of the InAs/InSb heterostructures  demonstrating sufficiently high recombination efficiency. In this work we take advantage of the recent progress in fabricating the narrow-bandgap InSb/InAs type-II quantum-confined heterostructures possessing bright PL caused by interband optical transitions \cite{Ivanov2005} in order to study the Zeeman effect and electron spin alignment by measuring spectra of circularly polarized PL in an external magnetic field in wide temperature- and excitation-level ranges.

The experimental samples contain ultrathin InSb insertions with a nominal thickness less than the Stranski-Krastanov critical value ($\sim$~1.7 monolayers (ML)), embedded into an InAs matrix. Such nanostructures demonstrate bright photo- \cite{Solov'ev2005} and electroluminescence \cite{Carrington2008} in the middle infrared spectral range up to room temperature. The InSb insertion can be considered as either a dense array of in-plane-coupled quantum dots \cite{Lyublinskaya2006} or an ultrathin disordered quantum well (QW). \cite{Terent'ev2009} Here we pursue the second approach and present the results of tight-binding calculations of the electron energy spectrum in the ultrathin InSb/InAs QW, comparing then these theoretical estimations with the experimental data of the magneto-optical studies. Nearly 100\% circular polarization of PL was observed at the magnetic field 3-4~T (depending on excitation power density $W_{exc}$) and temperatures lower than 5~K. The phenomenon is explained in terms of the complete electron and hole spin alignment, taking into account known selection rules for optical transitions. Revealed decreasing of the polarization degree with temperature or excitation density is generally explained in the model of repopulation of spin sublevels. The minor effect of polarization inversion in the case of high temperature or high excitation rate is described phenomenologically by introducing a parameter characterizing the dependence of the recombination efficiency on the level energy. The $z$-component of the heavy-hole g-factor is estimated from the experimentally obtained splitting of the PL band in the magnetic field  and the tabulated value of the electron g-factor for InAs.

\section{Samples and experimental technique}

The heterostructures with thin InSb insertions in InAs were fabricated by molecular beam epitaxy (MBE) using an approach exploiting As-Sb exchange reactions on the growth surface. The InSb layer was formed by an exposure of the InAs surface heated up to $T = 430\,^{\circ}\mathrm{C}$ to a Sb flux during 20 seconds. The maximum thickness of the InSb insertion grown by this method equals to 1~ML. Additional MBE deposition of InSb, following the InAs surface exposure, can be used to increase the insertion thickness. This two-stage procedure allows one to obtain the insertion thickness up to 2~ML. The details of the fabrication process are described elsewhere. \cite{Ivanov2005} The objects of the current studies are the structures with multiple 1-ML-thick InSb insertions separated from each other by 15-nm-thick InAs layers. The nominal thickness of the InSb insertion is derived from x-ray diffraction (XRD) studies. The active region of the structures was sandwiched between wider gap \nobreak{Al$_{0.2}$In$_{0.8}$As} layers needed to confine photoexcited carriers of both types within the region of the insertions.

One can assume that the growth of such ultrathin insertions may be accompanied by both intermixing and disordering. In our samples the effective length of these effects hardly exceeds 1-2 monolayers since the diffusion of atoms is extremely weak and does not surmount this value at such low temperature as $430\,^{\circ}\mathrm{C}$. It was shown that at temperatures lower then $490\,^{\circ}\mathrm{C}$ antimony atoms replace arsenic on a GaAs surface and the surface reconstruction revealed during the exposure of the GaAs surface to the antimony flux corresponds to 1 ML of GaSb. \cite{Whitman1999} A similar effect of replacing arsenic atoms by antimony was also observed in Refs \cite{Wang1993,Brown2002}. The opposite effect (As incorporation in Sb-based layer) is negligible at low temperatures but it should be taken into account at a higher growth temperature. \cite{Losurdo2006} One can expect a larger contribution emerging due to Sb segregation. It is known that MBE growth of antimonides on arsenides is accompanied by Sb and As intermixing in the growth direction due to the Sb segregation phenomenon. \cite{Steinshnider2000, Bennett1999} The effect of surface Sb segregation resulting in formation of a gradient InAsSb layer after overgrowing a thin InSb insertion by InAs was previously studied by us in detail. \cite{Semenov2007} We have shown that the Sb content in the first InAsSb monolayer overgrowing the InSb insertion at the growth temperature of $430\,^{\circ}\mathrm{C}$ doesn't exceed 20\%. As concerns the InSb/InAs structures studied in the paper, an original technique based on smart As soaking of growth surface was used, resulting in a strong (practically full) suppression of the Sb segregation as proved by RHEED (Reflection High-Energy Electron Diffraction) oscillation measurements (to be published elsewhere). Summarizing, our experimental heterostructures really contain a ML-scale InSb quantum well with probable weak disordering on the 1-2~ML scale. 

\begin{figure}
\includegraphics[width=0.5\textwidth]{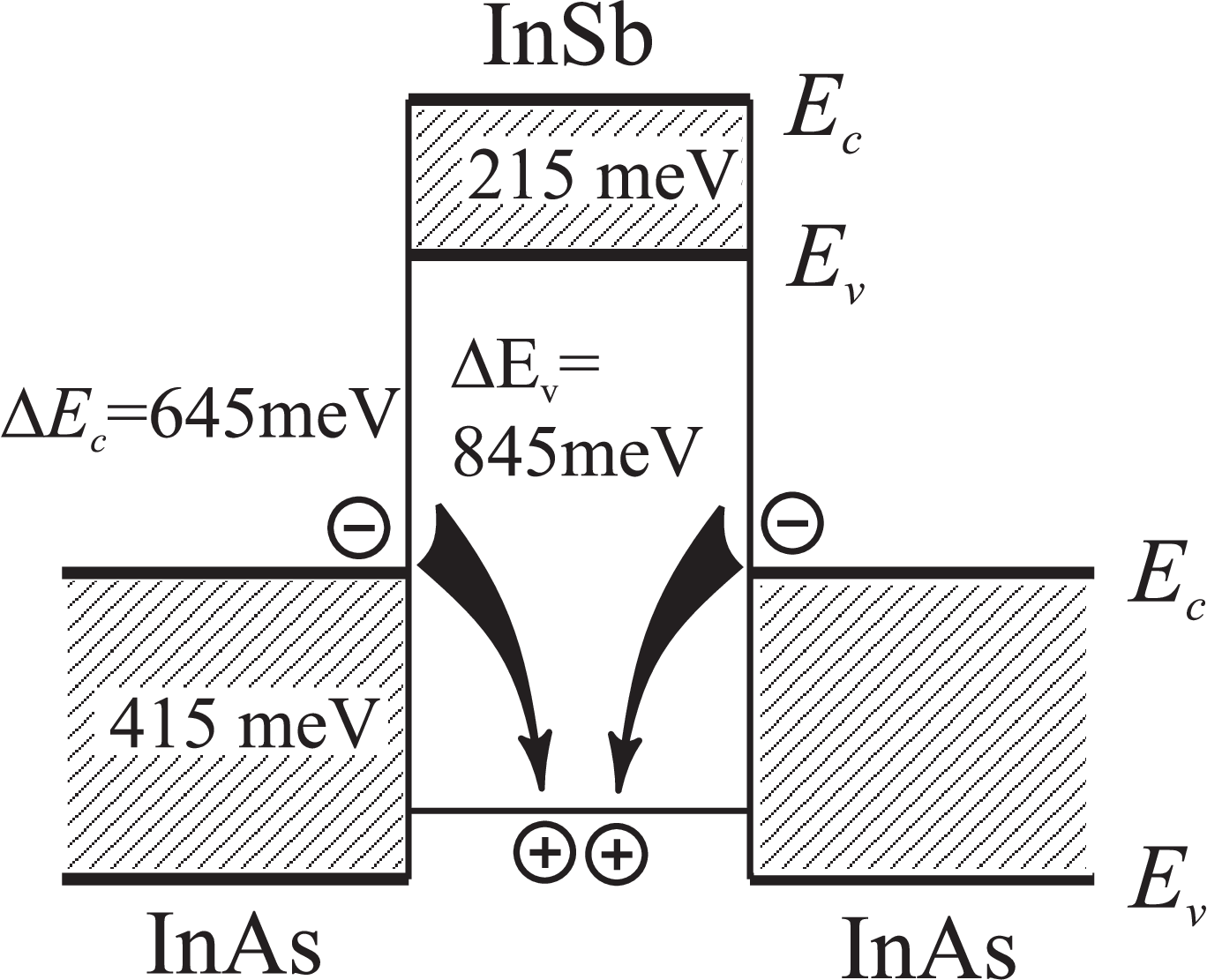}%
\caption{ \label{fig:band diagram} Band line-ups of a pseudomorphic InSb/InAs heterostructure calculated using a tight-binding approach. Indirect optical transitions involving holes confined in the InSb QW are shown by arrows.}
\end{figure}

\begin{figure}
\begin{minipage}{0.47\linewidth}
\includegraphics[width=\textwidth]{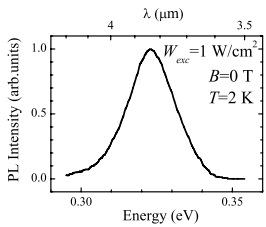}\\ a)%
\end{minipage}
\hfill
\begin{minipage}{0.47\linewidth}
\includegraphics[width=\textwidth]{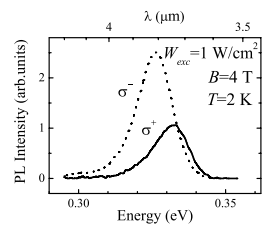}\\b)%
\end{minipage}
\caption{ \label{fig:PL} PL spectra taken at $T$ = 2~K and $W_{exc}$ = 1~W/cm$^{2}$. a) $B$ = 0~T, b) $B$ = 4~T.}
\end{figure}

This system is predicted to have a type-II broken-gap band alignment with the holes confined in the InSb QW. The schematic band diagram of the heterostructure of this type, calculated for a strained pseudomorphic InSb QW, is shown in Fig. \ref{fig:band diagram}. The electrons are located in the adjacent InAs layers; they are attracted to the holes via Coulomb interaction. Since the electrons and holes are spatially separated, only indirect in space optical transitions are possible. In this system the lowest optical transition energy is less than the InAs band gap and may be continuously varied in the range from 3.4 to 4.4~$\mu$m (at 80 K) by tuning the insertion width. \cite{Solov'ev2005} Another peculiarity of the system is a reduced overlap of the hole and electron wavefunctions that is expected to result in relatively weak oscillator strength of the optical transition and long radiative lifetime. Nevertheless the structures under study with the 1~ML thick InSb insertions demonstrate quite bright PL near 3.83~$\mu$m at zero magnetic field and $T$ = 2~K (see Fig.~\ref{fig:PL}a).

\begin{figure}
\includegraphics[width=0.5\textwidth]{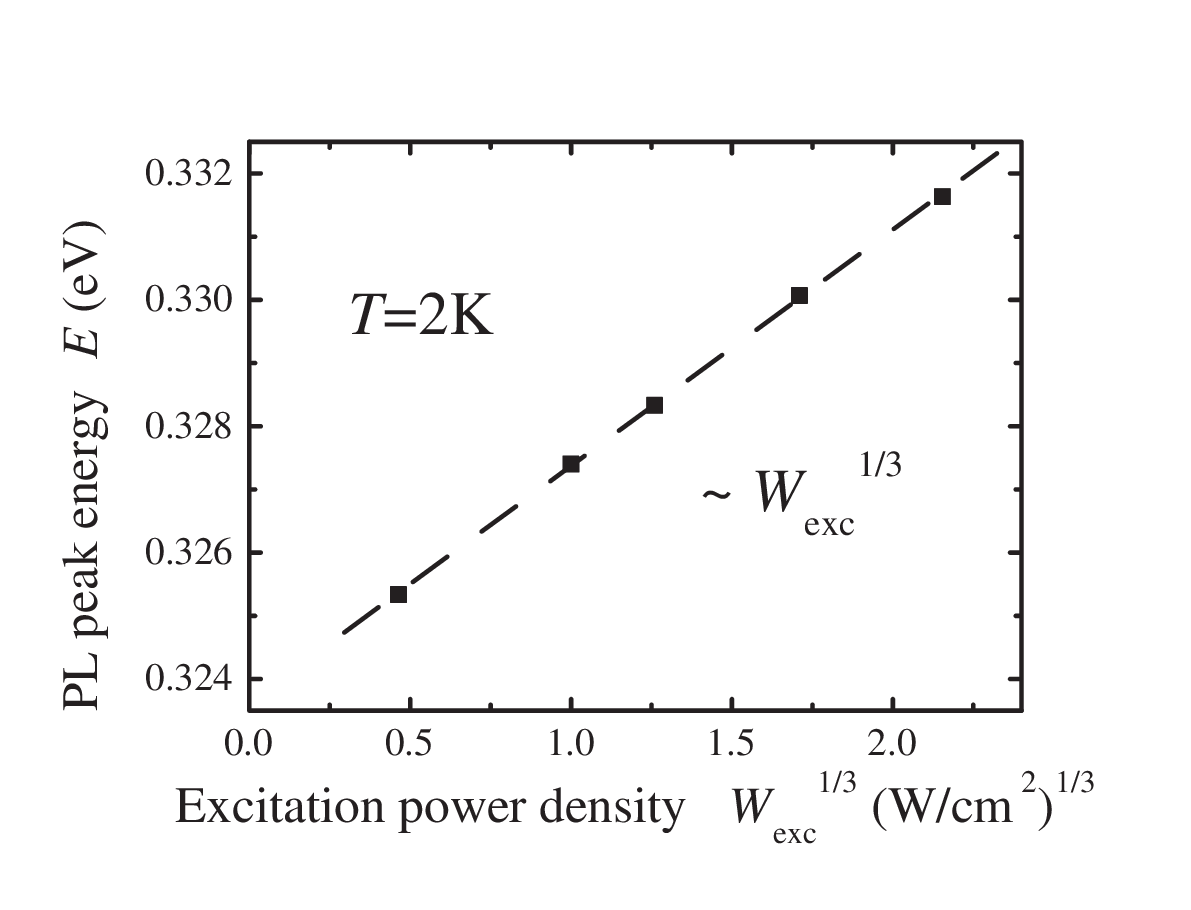}%
\caption{ \label{fig:E(I)} PL peak energy dependence on the cubic root of the excitation power density. The PL spectra of the InSb/InAs heterostructure were recorded at 2~K.}
\end{figure}

In contrast to type-I systems in type-II heterostructures the separation of the nonequilibrium electrons and holes in real space should result in the dipole layer and leads to formation of a quantum well for electrons in the vicinity of the interface. So the luminescence maximum shifts towards higher photon energies with rising excitation power density reflecting the increase in the electron quantization energy, which is proportional to $W_{exc}^{1/3}$. Hence, the emitting photon energy is expected to increase as the cubic root of the excitation power density. Such behavior is proved to be a characteristic feature of type-II systems. \cite{Ledentsov1995} Figure~\ref{fig:E(I)} shows the dependence of the PL peak energy on the excitation power density obtained for studied samples. The rising excitation density results in the predicted for type-II systems blue shift of the PL peak. These data provide an experimental proof that the  heterostructures based on strained InSb sheets in an InAs matrix have a type-II band alignment. 

The spectra of circularly polarized PL were measured in an external magnetic field applied in Faraday geometry. The experiments were carried out at magnetic fields $B$ up to 4~T in the temperature range from 2 to 120~K. PL was excited by a \textit{cw} laser diode operating at 809~nm. The laser emission was focused at the sample surface into a 1~mm size spot. An excitation density $W_{exc}$ was in the range of 0.03-40~W/cm$^{2}$. The PL circular polarization was analized by the combination of a NaCl Fresnel rhomb and a grating linear polarizer. The spectra were recorded by a grating monochromator equipped with a liquid nitrogen cooled InSb photovoltaic detector. A conventional lock-in technique was used to register the signal.

\section{Experimental results}
\subsection{Magnetic field dependence}
\begin{figure}
\begin{minipage}{0.47\linewidth}
\includegraphics[width=\textwidth]{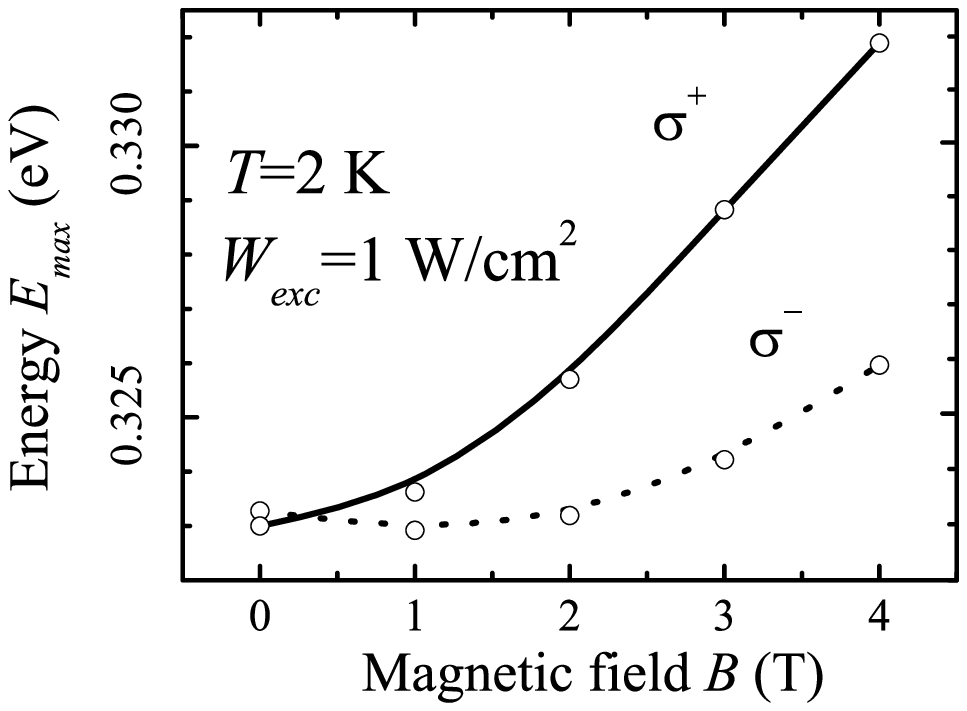}\\ a)%
\end{minipage}
\hfill
\begin{minipage}{0.47\linewidth}
\includegraphics[width=\textwidth]{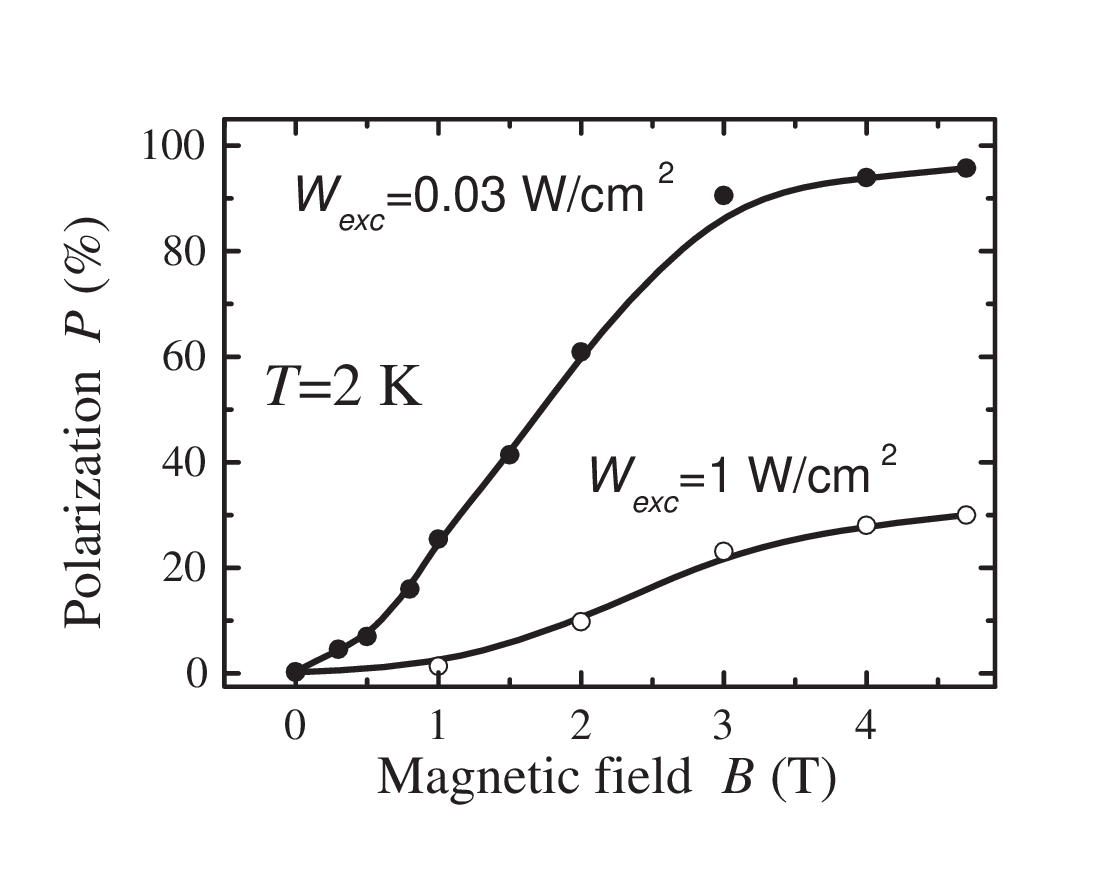}\\b)%
\end{minipage}
\caption{ \label{fig:f(B)} a) PL peak energy versus magnetic field at $T$ = 2~K and $W_{exc}$ = 1~W/cm$^{2}$. b) Magnetic field dependence of the polarization degree $P$ measured at the different excitation power densities.}
\end{figure}

Application of the external magnetic field drastically affects the PL spectrum. The PL peak splits into two circular-polarized components with different magnitude and shifts toward higher energies (see Fig.~\ref{fig:PL}b). The energy of the circularly polarized PL peaks $E_{max}$ and the polarization degree $P$ measured at a weak and intermediate power density are plotted in Fig.~\ref{fig:f(B)} as a function of the magnetic field ($P$ is defined as $P=\frac{I^{\sigma^{-}}-I^{\sigma^{+}}}{I^{\sigma^{-}}+I^{\sigma^{+}}}\cdot 100\%$, where $I^{\sigma^{+(-)}}$ is a  magnitude of the $\sigma^{+(-)}$ polarized peak). The magnetic field increase causes the polarization enhancement. At the excitation rate 1~W/cm$^{2}$, $P$ amounts to $\sim$~30\% at 4~T. At the same time at the weaker excitation it reaches almost 100\% even at 3~T (see Fig.~\ref{fig:f(B)}b). At magnetic fields $B$~\textless{}~2~T a diamagnetic shift of the PL lines is sublinear in contrast to moderate fields where the increase of $B$ results in a linear increase of the PL peak energy. Such behavior evidences the effect of magnetic freeze-out. This phenomenon will be briefly discussed afterwards, while in general it is out of scope of the present paper.

\begin{figure}
\includegraphics[width=0.5\textwidth]{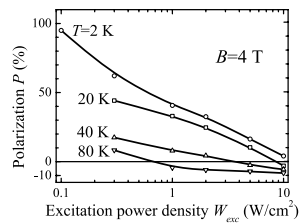}%
\caption{ \label{fig:f(I)} Polarization degree $P$ versus excitation power density $W_{exc}$ at different temperatures and $B$ = 4~T.}
\end{figure}

\subsection{Excitation power dependence}
We observed strong influence of the excitation power density $W_{exc}$ on the polarization degree (see Fig.~\ref{fig:f(I)}). At the lowest excitation rate almost 100\% $\sigma^{-}$ polarization was measured at low temperature. The polarization degree drops dramatically with the excitation intensity increase and changes its sign eventually. The inversion point moves to lower intensities when the sample temperature goes up. At the highest excitation intensity the polarization is $\sigma^{+}$ and approaches the value of 10\%.

\begin{figure}
\includegraphics[width=0.5\textwidth]{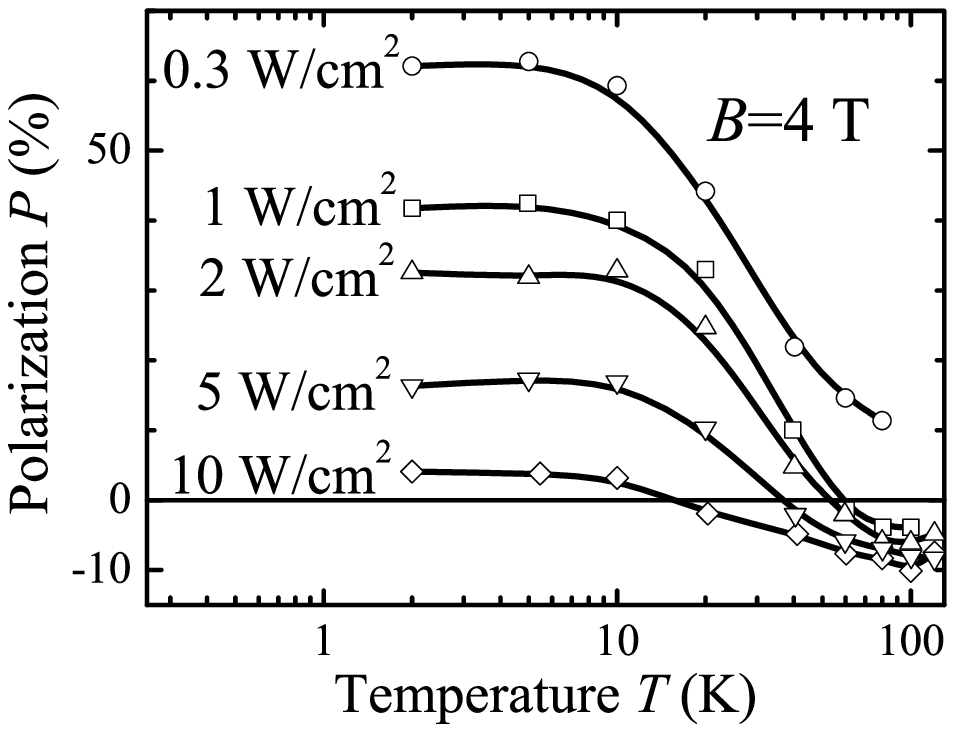}%
\caption{ \label{fig:f(T)} Temperature dependence of the polarization degree $P$ for different values of the excitation power density $W_{exc}$.}
\end{figure}

\subsection{Temperature dependence}
The temperature dependence of the polarization degree, measured at different excitation power densities is presented in Fig.~\ref{fig:f(T)}. The increase of the temperature results in a reduction of the polarization degree and inversion of its sign, resembling in general the excitation power dependence (Fig.~\ref{fig:f(I)}). The polarization degree saturates at the value $\sim$ 10\% $\sigma^{+}$ for high levels of excitation. At the temperature below $\sim$ 10~K the dependence is flat for any excitation rate.

\section{Discussion}
\begin{figure}
\includegraphics[width=0.5\textwidth]{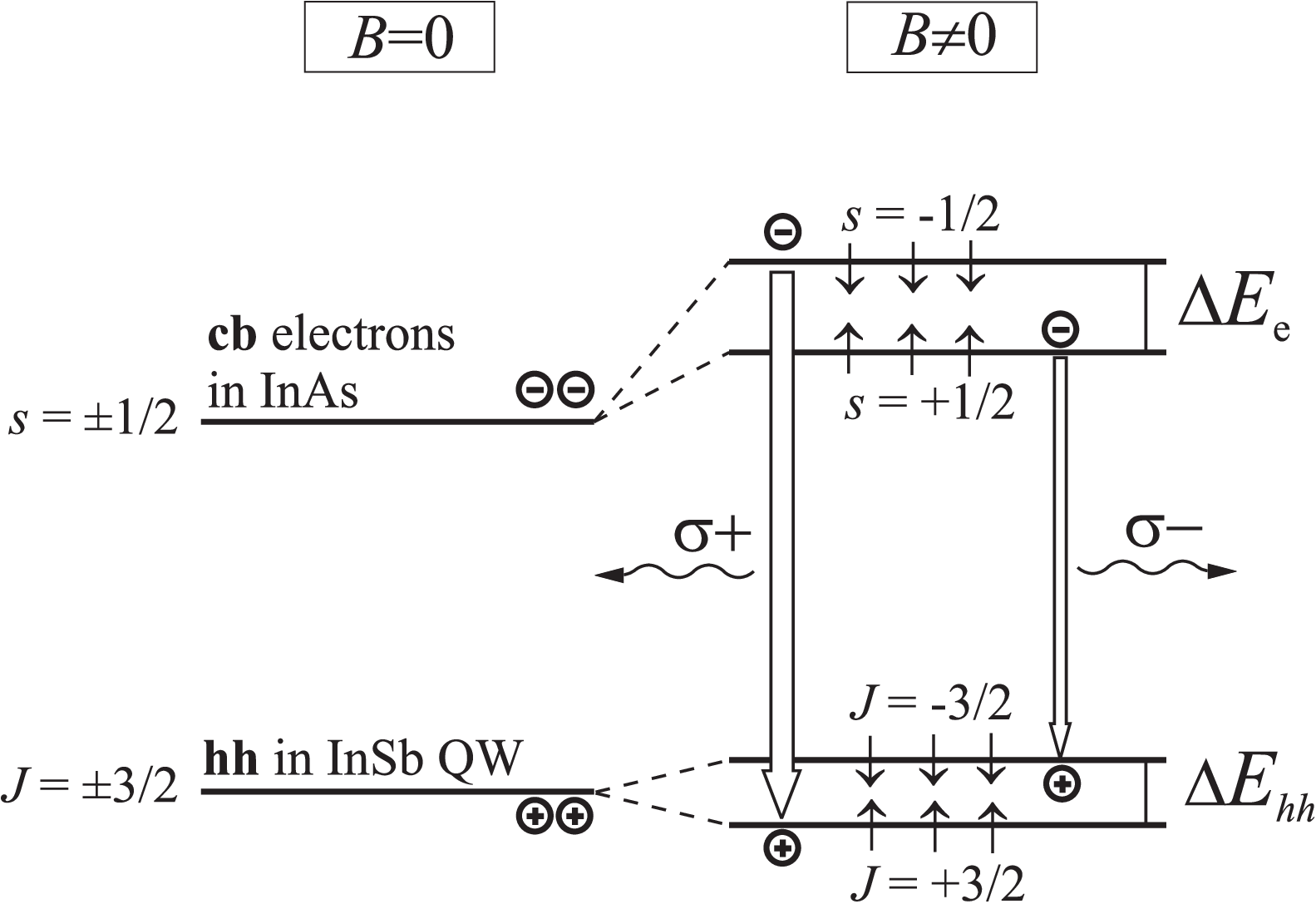}%
\caption{ \label{fig:selection rules} Spin splitting of electron and hole ground states in an external magnetic field in the InSb/InAs type-II QW along with allowed optical transitions. The larger thickness of the arrow showing optical transition indicates the higher recombination efficiency.}
\end{figure}

Optical transitions allowed in the considered structure under the magnetic field applied in Faraday geometry are shown schematically in Fig.~\ref{fig:selection rules}. As InSb and InAs form a type-II heterojunction, only indirect in space transitions involving holes confined in the InSb QW and electrons located in the surrounding InAs layers can occur. To define the sign of the polarization relevant to dominant optical transitions we have identified the type of holes occupying the lowest energy level in the QW and applied the selection rules shown in Fig.~\ref{fig:selection rules}. To find quantization levels in the QW a conventional approach based on the envelope function approximation (EFA) is not applicable here as (i) the InSb insertions are ultrathin ($\sim$~1~ML), (ii) employing EFA is questionable for the large ($\sim$~6\%) strain, and (iii) a complex interplay between quantum confinement and band mixing occurs in the heterostructure with broken-gap line-ups.

We have calculated the hole energy spectrum at a zero magnetic field using the tight-binding approximation which takes into account the full band structure of the system. In the tight-binding approach \cite{Slater1954} the wave function is written in the basis of atomic orbitals $\Phi_{\alpha}(\textbf r)$, which are assumed to be local:  $\psi(\textbf r)=\sum\limits_{n\alpha} C_{n\alpha}\Phi_{\alpha}(\textbf r-\textbf r_{n})$, where $n$ enumerates atoms and $\alpha$ enumerates basis functions at the atoms. We use the $sp^{3}d^{5}s^{*}$ model  where $\alpha$ runs through two $s$-like, three $p$-like and five $d$-like orbitals multiplied by basis spinors $\uparrow$ and $\downarrow$. \cite{Jancu1998} The Hamiltonian in this basis is reduced to a large sparse matrix with blocks corresponding to different atoms and chemical bonds between them. In the empirical variant of the method these blocks are filled by the parameters of bulk materials. Strain is taken into account by a certain change of parameters due to chemical bond lengths in the strained structure and some correction proportional to components of a strain tensor. \cite{Jancu2007} The tight-binding parameters chosen to fit band structure of both InAs and InSb are taken from most accurate \textit{ab initio} calculations.

The results of the computations performed for the InSb insertions of 1 and 2~ML thickness are presented in Table~\ref{tbl_hhlh}. The localization energy is measured from the top of the InAs valence band. In Table~\ref{tbl_hhlh}, we also provide wavelengths corresponding to the transitions between the given hole level and electrons at the bottom of the InAs conduction band. For larger thicknesses the calculations predict a relatively strong admixture of the light hole to heavy hole states. We also note that the levels position depends strongly on strain. Here we use a conventional elastic theory, which is proved to be adequate even for heterostructures with extremely thin layers.

{
\renewcommand{\arraystretch}{1.7} 
\setlength{\tabcolsep}{0.22cm}    
\begin{table}
\begin{tabular}{|c|c|c|c|}
\hline
$D$, ML
   & $E_{loc}$, meV & $\lambda$, $\mu$m
      &  Wave function  \cr \hline
\multirow{2}{*}{1}
   & $117$          & $4.13$
      &  $0.05\cdot \left\vert \pm\frac12 \right\rangle +
0.95\cdot\left\vert \pm\frac32\right\rangle$ \cr\cline{2-4}
   & $38$           & $3.27$
      &  $0.92\cdot \left\vert \pm\frac12 \right\rangle +
0.08\cdot\left\vert \pm\frac32\right\rangle$ \cr\hline
\multirow{2}{*}{2}
   & $297$          & $10.3$
      &  $0.07\cdot \left\vert \pm\frac12 \right\rangle +
0.93\cdot\left\vert \pm\frac32\right\rangle$\cr\cline{2-4}
   & $57$           & $3.44$
      &  $0.94\cdot \left\vert \pm\frac12 \right\rangle +
0.06\cdot\left\vert \pm\frac32\right\rangle$ \cr\hline
\end{tabular}
\caption{
   Energies and spin structure of the quantum-confined states in the InSb insertion of width $D$, calculated in the tight-binding approach. The localization energy $E_{loc}$ is measured from the top of the InAs valence band. For convenience, the wavelength $\lambda$ of the corresponding optical transition is also given. Last column contains a fraction of components with angular momenta 1/2 and 3/2 in the calculated wavefunctions.
}\label{tbl_hhlh}
\end{table}
}

The emission wavelength observed in the experiment with a reasonable accuracy corresponds to the value predicted for the hole ground states. However especially important is the identified dominant heavy hole origin of the hole ground state that enabled us to apply the respective selection rules (Fig.~\ref{fig:selection rules}). According to our calculations, the ground state of the QW is formed primarily by heavy holes with only 5\% admixture of light holes. Next level with considerably smaller localization energy is attributed to holes with the 1/2 spin projection, but the corresponding optical transition can not be observed in the PL experiment due to the small equilibrium population of carriers at relatively low temperatures. Thus comparison with the calculations allows one to conclude that the emission originates mainly from recombination of conduction band (cb) electrons in InAs and heavy holes (hh) localized in the InSb insertions.

If the external magnetic field is applied, Zeeman splitting occurs of both the electron and hole states. An electron g-factor in InAs is negative and equals to $\sim$~-15 in bulk material at the band bottom, so the ground electron spin level is occupied by carriers with the spin oriented along the magnetic field ($s=+1/2$). According to the selection rules, recombination of an electron at this state is allowed with a heavy hole with the projection of angular momentum on the magnetic field direction $J=-3/2$ (note that we use a hole representation for valence band states). This process is accompanied by emission of a $\sigma^{-}$ polarized photon. The optical transition involving electrons at the other spin state $s=-1/2$ with higher energy is possible only if an electron recombines with a heavy hole state with $J=+3/2$. It generates a $\sigma^{+}$ polarized photon. This model is consistent with the experimental data. Indeed the PL contour splits into two circularly polarized peaks if $B$ is applied. At a relatively low excitation level the lower energy peak is $\sigma^{-}$ polarized and has much higher magnitude than the other one.

The dependence of the optical transition energy $E$ on magnetic field can be expressed as a sum of the transition energy at zero magnetic field and the Landau and Zeeman splitting energies for both electrons in the InAs conduction band and heavy holes in the InSb insertions. 
\begin{equation}
\label{eq:Eg}
E(B) = E_{0} + \frac{1}{2} \hbar \omega_{c} \pm \frac{1}{2} g_{e} \mu_{B} B \pm \frac{3}{2} g_{hh} \mu_{B} B.
\end{equation}
Here $E_{0}$ is the optical transition energy at zero magnetic field, $\hbar \omega_{c}$ - electron cyclotron energy, $g_{e}$ - electron g-factor, $g_{hh}$ - heavy hole g-factor, and $\mu_{Â}$ - Bohr magneton. The term originating from the cyclotron energy of heavy holes is omitted here due to their relatively large effective mass. Note that the linear dependence $E(B)$ is modified in the weak-field region due to the effect of magnetic freeze-out. \cite{Yafet1956} It manifests itself as a sublinear growth of $E$ originating from a magnetic field induced increase of the carrier localization energy. This effect is essential in the range of $B$ where the electron cyclotron energy remains smaller than its localization energy and is well pronounced in narrow-gap semiconductors. In our case the magnetic freeze-out affects electrons localized in the shallow triangle QW emerging in the bulk InAs in the vicinity of the InSb insertions, in contrast to previous studies considering electrons localized by a shallow impurity potential. It is reasonable to expect that in our case electron localization energy is approximately equal to the shallow donor ionization energy in InAs which is a few meV. Taking this value as an estimation one can assess that the freeze-out contribution is essential at $B$~\textless{}~2~T that agrees with the experiment. To calculate $E(B)$ one should know g-factors of electrons and heavy holes. The g-factor of electrons in bulk InAs is known with high accuracy, but there are no experimental data on $g_{hh}$. Eq.~\ref{eq:Eg} fits well the experimental curves presented in Fig. \ref{fig:f(B)}a at $B$~\textgreater{}~2~T, if the $g_{hh}$ value is taken 3. We estimate the accuracy of this value as within 30\%. Thus our experiments enable one to estimate the heavy hole g-factor for the 1~ML thick InSb/InAs insertion for the carrier spin oriented normally to the QW plane.

At the relatively low excitation rate ($\sim$~0.1~W/cm$^{2}$) and low temperature (\textless{}~10~K), when only ground electron and hole states are populated, the observation of almost 100\% $\sigma^{-}$ PL polarization degree can be explained in terms of total spin alignment of electrons in the InAs conduction band and heavy holes in the InSb insertion due to Zeeman effect and the selection rules for optical transitions involving heavy holes. To explain the observed reduction of the polarization degree at the elevated temperature or excitation power density it is natural to take into consideration filling of higher-energy states. However the observed polarization inversion can hardly be explained in the framework of the simple model of thermal population. To conform the experimental data to the theory one should make a conjecture of different recombination efficiency of optical transitions involving different spin sublevels. This assumption seems reasonable for the type-II system where oscillator strength is determined by overlap of spatially separated electron and hole wave functions and, hence, strongly depends on the potential profile in the interface proximity that in turn is controlled by temperature and excitation rate. \cite{Hatami1998} Another factor to be taken into account can be recombination and spin dynamics of "bright" and "dark" excitons. \cite{Dyakonov2008} Accurate treatment of these phenomena is out of scope of this paper. Instead, we confine ourselves to the formalism using a phenomenological parameter $r$ that defines relative recombination efficiency of optical transitions involving electrons with spin "up" and spin "down". In the frame of the proposed model we have fitted the measured temperature dependence of the polarization degree $P$ by the following expression:

\begin{equation}
\label{P}
P = \frac{I^{\sigma^{-}} - I^{\sigma^{+}}}{I^{\sigma^{-}} + I^{\sigma^{+}}} = \frac{n\uparrow \cdot p\downarrow - r \cdot n\downarrow \cdot p\uparrow}{n\uparrow \cdot p\downarrow + r \cdot n\downarrow \cdot p\uparrow},
\end{equation}
where $I^{\sigma^{-}}$ and $I^{\sigma^{+}}$ are the magnitudes of  $\sigma^{-}$ and  $\sigma^{+}$ polarized PL peaks, respectively, $n\uparrow$ and $n\downarrow$ are the densities of electrons with $s=+1/2$ and $s=-1/2$, $p\uparrow$ and $p\downarrow$ are the densities of heavy holes with the angular momentum projection $+3/2$ and $-3/2$, and $r$ is a relative recombination efficiency for the optical transitions of electrons with different spin orientation. 
Under the condition of the low excitation limit, the population of electron states with spin "up" $f^{\uparrow}_{Be}$ may be approximated by the Boltzmann distribution 

\begin{equation}
\label{eq:Boltzman}
f^{\uparrow}_{Be} = e^{\frac{E_F - E_e^{\uparrow}}{kT}},
\end{equation}
where $E_{F}$ is the Fermi energy and $E_{e}^{\uparrow}$ is the electron level energy with spin "up".
The density of electrons with spin "up" $n\uparrow$ is given by Eq.~\ref{eq:concentration}, where $N_{e}^{\uparrow}$ is an effective density of states:

\begin{equation}
\label{eq:concentration}
n\uparrow = N^{\uparrow}_{e} \cdot f^{\uparrow}_{Be}.
\end{equation}

\begin{figure}
\includegraphics[width=0.5\textwidth]{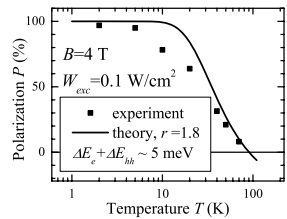}%
\caption{ \label{fig:fitting} Experimental and simulated dependences of the polarization degree versus temperature. The best fit is obtained if the relative oscillator strength $r$ equals to 1.8.}
\end{figure}

Similar expressions can be written for the other electron spin projection and for holes. After routine manipulations one can derive the equation for the polarization degree as

\begin{equation}
\label{eq:polarization}
P = \tanh\left[\frac{\Delta E_{e} + \Delta E_{hh}}{2kT} - \frac{\ln(r)}{2} \right],
\end{equation}
where $\Delta E_{e} = E_{e}^{\downarrow} - E_{e}^{\uparrow}$ is Zeeman splitting of the InAs conduction band, $\Delta E_{hh} = E_{hh}^{\uparrow} - E_{hh}^{\downarrow}$ is Zeeman splitting of the heavy holes localized in the InSb insertions, $r$ is the relative recombination efficiency for the optical transitions of electrons at different spin states. This expression is valid at the assumption of the Boltzman carrier distribution and can be employed only in the limit of the lowest excitation rate.
Figure~\ref{fig:fitting} shows the experimental temperature dependence of the polarization degree (squares) and the results of the simulation performed according to Eq.~\ref{eq:polarization} (solid curve). The value $\Delta E_{e} + \Delta E_{hh}$ was taken as splitting of the circular polarized PL lines. The experimental data shown in Fig.~\ref{fig:fitting} were measured at the excitation power density 0.1~W/cm$^{2}$ at the magnetic field 4~T. Within the experimental accuracy the PL peak splitting is independent of the excitation power density, being approximately equal to 5~meV at the magnetic field 4~T. The best fitting of the experimental data is obtained at $r = 1.8$.

\begin{figure}
\includegraphics[width=0.5\textwidth]{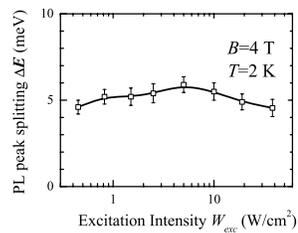}%
\caption{ \label{fig:splitting} PL peak splitting versus excitation power density at $T$ = 2~K and $B$ = 4~T.}
\end{figure}

Another approach to the estimation of $r$ can be employed, if one notices that in the limit of high temperatures Eq.\ref{eq:polarization} transforms to
\begin{equation}
\label{eq:simple}
P_{T \to \infty} = \frac{1 - r}{1 + r}.
\end{equation}

Then
\begin{equation}
\label{eq:strength}
r = \frac{1 - P_{T \to \infty}}{1 + P_{T \to \infty}}.
\end{equation}

Figure~\ref{fig:f(T)} indicates that the polarization degree saturates at the value $\sim$~10\% $\sigma^{+}$ in the case of high temperature. We have evaluated $r$ by substituting this value into Eq.~\ref{eq:strength}. The obtained estimation ($r$~\textgreater{}~1.2) does not contradict to the magnitude found by fitting the experimental data with Eq.~\ref{eq:polarization}. In the framework of the model, the reduction of $P$ versus temperature is caused by the change of the relative population of electron and heavy hole spin sublevels. Therefore the increase in excitation density should affect $P$ in the same manner. This deduction correlates well with Fig.~\ref{fig:f(I)} showing some decrease of $P$ when the excitation level increases.

\section{Conclusion}
Magneto-PL of type-II narrow-bandgap heterostructures with an ultrathin InSb QW in an InAs matrix has been studied in Faraday geometry. Circularly polarized spectra have been measured in a wide range of excitation intensity at temperatures varied from 2 to 120~K. The emission is almost 100\% $\sigma^{-}$ polarized at the conditions of the lowest excitation rate and temperature even at a moderate magnetic field. That evidences strong electron and hole spin alignment due to Zeeman splitting of electron and heavy hole states in the InSb/InAs heterostructure. The sign of polarization conforms to the selection rules for optical transitions involving heavy holes. To elucidate the origin of the hole ground state in the InSb QW a calculation of the energy spectrum was performed using the tight-binding approximation taking into account the full band structure of the system. It was found that the hole ground state is formed primarily by heavy holes with a little admixture of light holes, that is consistent with $\sigma^{-}$ light polarization observed in experiment. An increase of either temperature or excitation intensity results in a reduction of the polarization degree and change of its sign. The effect is explained in terms of either thermal or non-equilibrium optical population of the upper electron and hole spin sublevels. The polarization sign inversion occurs due to larger recombination efficiency for the upper level at the condition of approximately equal population of the spin sublevels. The heavy hole g-factor for the 1~ML thick InSb QW and spin direction normal to the QW plane has been experimentally estimated as 3.

The principal opportunity of using narrow-band semiconductors with large intrinsic electron g-factor for implementation of the spin initiation function has been demonstrated by the example of the InSb/InAs heterostructures. Rather strong magnetic fields and relatively low temperatures that are necessary to create 100\% spin polarization in this sample may be appreciably improved by designing a heterostructure where electrons are localized in InSb having much larger electron g-factor than InAs.

\begin{acknowledgments}
The financial support of the work by RFBR, Russian Ministry of Education and Science and project FP7 ITN SPIN-OPTRONICS is gratefully acknowledged. The authors thank L.E. Golub (the Ioffe Institute) for helpful discussion. 
\end{acknowledgments}

\end{document}